\documentclass{elsart}
\usepackage{graphicx}
\usepackage{amsmath}
\usepackage{amsfonts}
\usepackage{amssymb}
\begin{document}
\begin{frontmatter}
\title{A Large N expansion for Gravity}
\author{F. Canfora}
\address
{Istituto Nazionale di Fisica Nucleare, Sezione di Napoli, GC di Salerno\\
Dipartimento di Fisica ''E.R.Caianiello'', Universit\`{a} di
Salerno\\ Via S.Allende, 84081 Baronissi (Salerno), Italy}
\ead{canfora@sa.infn.it}

\begin{abstract}
A Large N expansion for gravity is proposed. The scheme is based
on the splitting of the Einstein-Hilbert action into the BF
topological action plus a constraint. The method also allows to
include matter fields. The relation between matter and non
orientable fat graphs in the expansion is stressed; the special
role of scalars is shortly discussed. The connections with the
Holographic Principle and higher spin fields are analyzed.
\end{abstract}
\end{frontmatter}

\section{Introduction}

\noindent Large \textbf{N}-expansion for a \textbf{SU(N)}\ \textit{Gauge
Theory}, introduced by 't Hooft in \cite{T74a} \cite{T74b}, is indeed one of
the most powerful non perturbative techniques available to investigate non
linear gauge theories. In particular, even if large \textbf{N} \textbf{SU(N)}%
\ Gauge Theory has not been solved yet, the large \textbf{N}-
expansion provides the issues of \textit{confinement},
\textit{chiral symmetry breaking} and the relation with \textit{string theory%
} with a rather detailed understanding. The Veneziano limit \cite{Ve76},
which corresponds to a 't Hooft limit in which the ratio \textbf{N/N}$_{f}$
is kept fixed (\textbf{N}$_{f}$ being the number of quarks\ flavours), shed
further light on non perturbative features of large \textbf{N} \textbf{SU(N)}%
\ Gauge Theory clarifying several features of quarks and mesons
dynamics; while a clear analysis of the role of Baryons at large
\textbf{N} (which, in many respects, behave as solitons) has been
provided in \cite{Wi79} (see, for two detailed pedagogical
reviews, \cite{Ma98}). It is also worth to mention that (several
types of) large \textbf{N}-expansions played a prominent role to
provide the Holographic Principle with further supports and
practical realizations (see, for a detailed review, \cite{AG00a}).
For the above reasons, it would be very important to have at one's
disposal similar techniques to investigate the non perturbative
features of Einstein-Hilbert action. Indeed, \textit{General
Relativity} bears a strong resemblance with Gauge Theory because
of the several common geometrical structures (such as the
curvature tensor "measuring" the deviation from flatness, the
"minimal coupling rule" which allows to rewrite Lorentz invariant
equations in the presence of a non trivial gravitational field by
substituting ordinary derivatives with covariant derivatives and
so on). On the other hand, there are many striking differences
which make General Relativity and Gauge Theory very unlike
theories: first of all, while Gauge Theory is a theory defined on
a fixed background space-time, General Relativity determines the
dynamics of the background spacetime itself; the actions of
General Relativity and Gauge Theory are different and General
Relativity is not perturbatively renormalizable in the ordinary
field theoretical sense. There are many others technical
differences, here it is worth to mention that in Gauge Theory
there is a clear separation between internal and space-time
symmetries (and this allows to consider the large
\textbf{N}-expansion by increasing the internal symmetry group
while keeping fixed the space-time structure), for the above
reasons in General Relativity this is not the case and it is not
obvious what a large \textbf{N}-expansion in gravity could be.

An interesting attempt to obtain a similar expansion in gravity
has been performed in \cite{To77} and \cite{Str81} and further
refined in \cite{Bo03}: in these papers the authors argued that
the role of the internal index \textbf{N} on the Gauge Theory's
side could be played on the General Relativity's side by the
number of space-time dimensions: in other words, the authors
proposed an expansion in which the small parameter is $1/D$ where
$D$ is the number of space-time dimensions. The result of their
analysis is that in a large $D$
expansion of the Einstein-Hilbert action, unlike the large \textbf{N}%
-expansion of Gauge Theory in which all the planar diagrams are on
equal footing, a subclass of the full set of planar diagrams
dominates. Unfortunately, their analysis is affected by some
technical problems which are absent in the Gauge Theory case. In
particular, in a large $D$ expansion, the power of $D$ in a given
diagram of the expansion is determined not only by the topology of
the two dimensional surface on which such a diagram can be drawn
(as it happens in the Gauge Theory case): a prominent role is also
played by space-time and/or momentum integrals which also depend
on $D$ and
a complete analysis of the contribution of the above integrals to the $D-$%
dependence of any diagram is a rather hopeless task \cite{Str81} \cite{Bo03}%
. Another technical problem related to the previous one is the
fact that, in a large $D$ expansion, it is necessary to consider
gravitational fluctuations around the trivial flat solution: the
reason is that it is not possible to disentangle the space-time
dependence due to space-time and/or momentum integrals from the
"internal indices" dependence of any diagram since these two
different kinds of structures overlap in the General Relativity
case. If it would be possible to single out suitable "internal
indices" in General Relativity then one could achieve an expansion
whose topological character can be analyzed independently on the
background: a non trivial background would only affect the
space-time and/or momentum integrals of the diagrams while the
internal index structure (which plays the prominent role in the
large \textbf{N} limit) would not change.

Here a scheme is proposed to overcome the above difficulties of
the large $D$ expansion which is based on the
\textit{BF}-theoretical formulation of Einstein-Hilbert action in
which the General Relativity action is splitted into a topological
term plus a constraint (the analysis of the relations between
General Relativity and \textit{BF} theory has a long history: see,
for example, \cite{Ca91} \cite{Ca91b} \cite{De99} \cite{Re99}
\cite{Ca01} and references therein): this way of writing the
General Relativity action allows to distinguish internal index
structure from space-time index structure and is strictly related
to the connection formulation(s) of General Relativity which
played a prominent role in the formulation of \textit{Loop Quantum
Gravity} (for a detailed review see, for example, \cite{As04}).
Moreover, it is possible to make an interesting comparison between
the BF formulation of General Relativity and Yang-Mills theory
introduced in \cite{FTZ96}.

The paper is organized as follows: in the second section the
formulation of General Relativity as a constrained BF theory is
shortly described. In the third section a suitable
\textquotedblright large \textbf{N} 't Hooft
expansion\textquotedblright\ on the internal indices is
introduced. In the fourth section the inclusion of matter and a
\textquotedblright Veneziano-like\textquotedblright\ limit are
discussed. In the fifth section, the General Relativity and the
Gauge Theory expansions are compared and the connections with the
Holographic Principle and higher spin fields are analyzed.
Eventually, the conclusions are drawn.

\section{BF-theory and gravity}

In this section the formulation of General Relativity as a
constrained \textit{BF} theory will be shortly described along the
lines of \cite{Ca91} \cite{Ca91b} \cite{De99} \cite{Re99}
\cite{Ca01}.

\textit{BF} theories are topological theories, that is, they exhibit the
following properties: firstly, they are defined without any reference to a
background metric and, moreover, they have no local degrees of freedom.
Here, only the four dimensional case will be considered. The \textit{BF}
theory in four dimensions is defined by the following action%
\begin{align}
S\left[  A,B\right]   &  =%
{\displaystyle\int\limits_{M}}
B^{IJ}\wedge F_{IJ}(A)=\frac{1}{4}%
{\displaystyle\int\limits_{M}}
\varepsilon^{\alpha\beta\gamma\delta}B_{\alpha\beta}^{IJ}F_{\gamma\delta
IJ}d^{4}x\label{bfa}\\
B^{IJ}  &  =\frac{1}{2}B_{\alpha\beta}^{IJ}dx^{\alpha}\wedge
dx^{\beta},\quad
F_{IJ}=\frac{1}{2}F_{\alpha\beta IJ}dx^{\alpha}\wedge dx^{\beta}\nonumber\\
F_{\alpha\beta IJ}  &  =\left(
\partial_{\alpha}A_{\beta}-\partial_{\alpha }A_{\beta}\right)
_{IJ}+A_{\alpha I}^{L}A_{\beta LJ}-A_{\beta I}^{L}A_{\alpha
LJ}, \label{fieldstr}
\end{align}
where $M$ is the four-dimensional space-time, the greek letters
denote space-times indices, $\varepsilon ^{\alpha \beta \gamma
\delta }$\ is the
totally skew-symmetric Levi-Civita symbol in four-dimensional space-times, $I$%
, $J$ and $K$ are Lorentz (internal) indices which are raised and lowered
with the Minkowski metric $\eta _{IJ}$: $I,J=1,..,D$. Thus, the basic fields
are a $so(D-1,1)$-valued two form $B_{IJ}$ and a $so(D-1,1)$ connection one
form $A_{\alpha LJ}$, the internal gauge group being $SO(D-1,1)$. Also the
Riemannian theory can be considered in which the internal gauge group is $%
SO(D)$ and the internal indices are raised and lowered with the euclidean
metric $\delta _{IJ}$; in any case, both $B_{IJ}$ and $A_{\alpha LJ}$ are in
the adjoint representation of the (algebra of the) internal gauge group:
this simple observation will be important in order to develop a 't Hooft
like large \textbf{N} expansion. The equations of motion are
\begin{equation}
F=0,\quad \nabla _{A}B=0  \label{vabf}
\end{equation}
where $\nabla _{A}$ is the covariant derivative with respect to
the connection $A_{\alpha LJ}$. The above equations tell that
$A_{\alpha LJ}$ is, locally, a pure gauge and $B^{IJ}$\ is
covariantly constant. Obviously, the \textit{BF }action does not
describe the dynamics of general relativity which, indeed, has
local degrees of freedom. On the other hand, if $B^{IJ}$ would
have this form

\begin{equation}
B^{IJ}=\frac{1}{2}\varepsilon _{KL}^{IJ}e^{K}\wedge e^{L}  \label{preco}
\end{equation}

then the action (\ref{bfa}) would be nothing but the Palatini form
of the generalized Einstein-Hilbert action (obviously, standard
General Relativity is recovered when $D=4$). It turns out that eq.
(\ref{preco}) can be enforced by adding to the action (\ref{bfa})
a suitable constraint; thus, the basic action in the BF formalism
is

\begin{equation}
\kappa S_{GR}=S\left[  A,B\right]  -%
{\displaystyle\int\limits_{M}}
\left(  \phi_{IJKL}B^{IJ}\wedge B^{KL}+\mu H\left(  \phi\right)
\right)
\label{gra}%
\end{equation}

where $\kappa $ is the gravitational coupling constant, $\mu $ is a
four-form and $H(\phi )$ is a scalar which can have the following three
expressions:%
\begin{equation}
H_{1}=\phi _{IJ}^{IJ},\quad H_{2}=\phi _{IJKL}\varepsilon ^{IJKL},\quad
H_{3}=a_{1}H_{1}+a_{2}H_{2},  \label{enfo1}
\end{equation}%
where $a_{i}$ are real constants. It is worth to note here that the scalar $%
\phi $ takes value in the tensor product of the adjoint representation of $%
so(D-1,1)$ (or of $so(D)$) with itself. The form (\ref{gra}) of the
Einstein-Hilbert action will be the starting point of the \textquotedblright
gravitational\textquotedblright\ large \textbf{N} expansion.

\section{A gravitational large \textbf{N} expansion}

In this section, a\ 't Hooft like expansion based on the form (\ref{gra}) of
the Einstein-Hilbert action is proposed: this scheme allows an interesting
comparison with the BF formulation of Yang-Mills theory introduced in \cite%
{FTZ96}.

The first question which should be answered in order to set in a suitable
framework for a gravitational large \textbf{N} expansion is: who is \textbf{N%
} in the gravitational case? The suggestion of \cite{Str81}
\cite{Bo03} is that \textbf{N} should be identified with the
number of space-time dimensions $D$. Indeed, the basic variables
of the \textit{BF}-like formulation (and, in general, of any
connection formulation of General Relativity) have the great merit
to manifest a natural separation between space-time and internal
indices. It becomes possible to consider a limit in which the
space-time dimensionality is fixed while the dimension of the
internal symmetry group approaches to infinity. In the Gauge
Theory case, the gluonic fields are in the adjoint of
\textbf{U(N)} (which can be thought as the tensor product of the
fundamental and the anti-fundamental representations) while the
quarks in the fundamental. It is then possible to introduce the
celebrated 't Hooft double line notation in which it is easy to
show that, in a generic diagrams, to every closed color line
corresponds a factor of \textbf{N} which is nothing but the
dimension of the fundamental representation of \textbf{U(N)}.

In the General Relativity case, two of the basic fields ($A$ and
$B$) are $p$-forms taking values in the adjoint representation of
$so(D-1,1)$: in other words, in the internal space, $A$ and $B$
are real $D\times D$ skew-symmetric matrices carrying, therefore,
two vectorial indices running from $1$ to $D$. While the
Lagrangian multiplier $\phi $ carries four vectorial indices
running from $1$ to $D$ (this fact will play a role in the
following). The
above considerations clarify that, in the gravitational case, \textbf{N}$%
^{-1}$ should be identified with $1/D$:%
\begin{equation}
\mathbf{N}^{-1}=D^{-1};  \label{seitu}
\end{equation}%
however, in order to avoid confusion with the notation of \cite{Str81} \cite%
{Bo03} and with the letter which is often used to denote the
number of space-time dimensions (which, in fact, in this approach
is kept fixed), the left hand side of eq. (\ref{seitu}) will be
used to denote the expansion parameter. An important benefit of
the connection formulation is that the double line notation can be
fairly adopted: the only difference with respect to
the Gauge Theory case is that, being the fundamental representation of $so(%
\mathbf{N}-1,1)$ real, the lines of internal indices carry no arrow\footnote{%
The point is that in the Gauge Theory case one has to distinguish
the fundamental and the anti-fundamental representations of (the
algebra of) \textbf{U(N):} this is achieved by adding to the color
lines incoming or outgoing arrows.}.

\begin{figure}
\begin{center}
\includegraphics*[scale=.40]{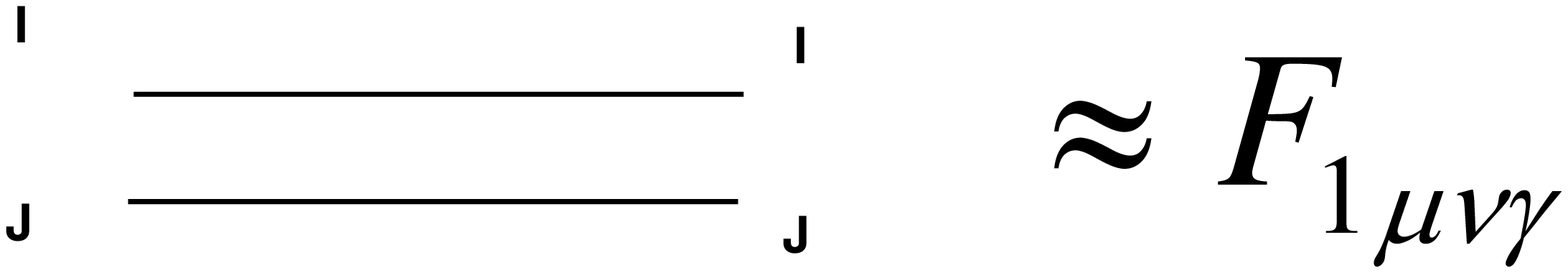}
\end{center}
\caption{As in standard gauge theories, the structure of the
propagator implies that internal indices are conserved along the
internal lines.} \label{Pprop}
\end{figure}

In order to provide the large \textbf{N} expansion with the usual
topological classification one should write down the Feynman rules
for the\ Einstein-Hilbert action (\ref{gra}) plus the gauge-fixing
and ghost terms. As far as a large \textbf{N} expansion is
concerned, the ghost terms are not important since they do not
influence the topological character of the expansion and the
topological classification of the diagrams (see, for example,
\cite{Ma98}). In this case it is convenient to find the Feynman
rules along the lines of \cite{Mar97} where the authors considered
the case of the \textit{BF} formulation of Yang-Mills theory: in
this way it will be easier to make a comparison between the large
\textbf{N} expansion in General Relativity and Gauge Theory. Thus,
the starting point is the action

\begin{equation}
S_{GR}=\frac{1}{\kappa}\left[  S\left[  A,B\right]  -\frac{c_{2}}{2}%
{\displaystyle\int\limits_{M}}
\left(  \phi_{IJKL}B^{IJ}\wedge B^{KL}+\mu H\left(  \phi\right)
\right)
\right]  \label{spa}%
\end{equation}

where $\kappa $\ is the gravitational coupling constant, the real constant $%
c_{2}$ keeps track of the terms which distinguish
\textit{BF}-theory from General Relativity. The natural choice is
to consider as the Gaussian part the off-diagonal kinetic term

\begin{equation}
S_{0}=\frac{1}{\kappa}%
{\displaystyle\int\limits_{M}}
\left(  \varepsilon^{\alpha\beta\gamma\delta}B_{\alpha\beta}^{IJ}%
\partial_{\gamma}A_{\delta IJ}\right)  , \label{prop1}%
\end{equation}

in such a way that the $A\rightarrow B$ propagator (which
propagates $A_{\mu }$ into $B_{\nu \gamma }$) has the following
structure (see fig. (\ref{Pprop})):
\begin{align}
\Delta _{(A,B)\mu \nu \gamma }^{(IJ,KL)} &=\delta ^{IL}\delta
^{JK}F_{1}(p)_{\mu \nu \gamma } \\
F_{1}(p)_{\mu \nu \gamma } &= -\frac{1}{2}\varepsilon _{\mu \nu
\gamma \alpha }\frac{p^{\alpha }}{p^{2}} \label{propag1}
\end{align}
where $F_{1}(p)_{\mu \nu \gamma }$ (which is the
space-time-momentum dependent part of the
propagator\footnote{$F_{1}(p)_{\mu \nu \gamma }$ has a form
similar to the $A\rightarrow B$ propagator in \cite{Mar97} in the
usual
Feynman gauge; the procedure to find the explicit expression of $%
F_{1}(p)_{\mu \nu \gamma }$ is analogous to the procedure of
\cite{Mar97} and \cite{Cat98} but, as far as the purposes of the
present paper are concerned, is not relevant since here only the
internal index structure is needed.}) is not relevant as far as
the large \textbf{N} expansion is concerned. The internal index
structures of the $A\rightarrow A$ propagator and of the
$B\rightarrow
B$ propagator are analogous to the expressions found in \cite{Mar97}:%
\begin{align}
\Delta _{(A,A)\mu \nu }^{(IJ,KL)} &=\delta ^{IL}\delta
^{JK}F_{2}(p)_{\mu \nu },\quad \Delta _{(B,B)\mu \nu \gamma \rho
}^{(IJ,KL)}=\delta ^{IL}\delta ^{JK}F_{3}(p)_{\mu \nu \gamma \rho
} \\
F_{2}(p)_{\mu \nu } &=\frac{1}{p^{2}}(\delta ^{\mu \nu
}-\frac{p^{\mu }p^{\nu }}{p^{2}})+\alpha _{1}\frac{p^{\mu }p^{\nu
}}{p^{4}} \label{propag2} \\
F_{3}(p)_{\mu \nu \gamma \rho } &=-\alpha _{2}\frac{p^{\mu }p^{\nu
}}{p^{4}} \label{propag3}
\end{align}%
where $F_{2}$ and $F_{3}$\footnote{%
A procedure to deduce them can be found, for example, in
\cite{Mar97} and \cite{Cat98}.} (which are the space-time-momentum
dependent parts of the propagators) will not be relevant as far as
the large \textbf{N} expansion is concerned and $\alpha _{1}$ and
$\alpha _{2}$ are real gauge parameters. Actually, even if $F_{2}$
and/or $F_{3}$ would vanish, the topological classification of the
fat graphs and the large \textbf{N} expansion would not change: it
is only important to note that, in a double-line notation, the
internal index is conserved along the \textquotedblright
color\textquotedblright\ lines.

A remark is in order here: in the BF Yang-Mills case the action
can be written as follows

\begin{equation}
S_{BFYM}=S\left[  A,B\right]  -g^{2}%
{\displaystyle\int\limits_{M}}
B^{IJ}\wedge\ast B_{IJ} \label{bfym1}%
\end{equation}

where $\ast $ is the Hodge dual. The second term on the right hand
side can be considered both as a part of the kinetic Lagrangian
and as a true vertex. The first choice gives rise to standard
quantization procedure via gauge fixing and ghost terms. The
second choice instead would be rather involved because one should
gauge fix\ also the topological symmetry of the BF action and this
would require a \textquotedblright ghost of
ghost\textquotedblright\ structure. In fact, in the General
Relativity case, there is no choice, the only available
possibility is the (more involved) second
one: in this case the quantization procedure to follow can be found in \cite%
{Cat98}. The point is that the quadratic term of the BF Yang-Mills case is
replaced by a (rather unusual) cubic interaction term. However, as it has
been already remarked, from a large \textbf{N} perspective the ghost terms
are not relevant\footnote{%
The reason is that, from an "internal lines" perspective, they add no new
vertices: in other words, the ghost vertices have the same internal index
structures and coupling constants of the physical vertices (see, for
example, \cite{Ma98}).}.

\begin{figure}[h]
\begin{center}
\includegraphics*[scale=.50]{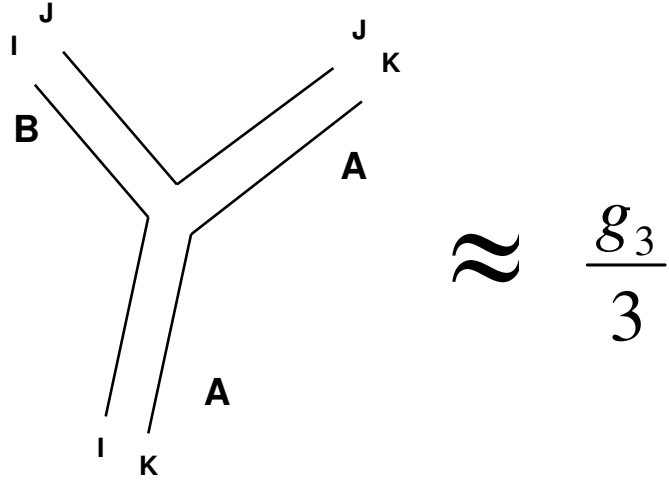}
\end{center}
\caption{This vertex is similar to the BF Yang-Mills vertex.}
\label{Vvert2}
\end{figure}

Unlike the already mentioned Yang-Mills case, the theory has two vertices:%
\begin{equation}
V_{1}(A_{\mu}^{a},A_{\nu}^{b},B_{\alpha\beta}^{c})=\frac{g_{3}}{3}%
f^{abc}\varepsilon_{\mu\nu\alpha\beta},\quad
V_{2}(B_{\mu\nu}^{a},B_{\alpha\beta}^{b},\phi^{cd})=\frac{c_{2}}{2}%
\delta_{ac}\delta _{bd}\varepsilon^{\mu\nu\alpha\beta}   \label{vertex}
\end{equation}
where $a$, $b$, $c$ and $d$ are internal indices in the adjoint
representation, $f^{abc}$ are the structure constants, $g_{3}$ (which can be
assumed to be positive) and $c_{2}$ are adimensional coupling constants
which keep track of the vertices in the large \textbf{N} counting and the
reason of the seemingly strange normalization of $g_{3}$ and $c_{2}$ in the
above equation will be clarified in the next subsection (see fig. (\ref%
{Vvert1}) and fig. (\ref{Vvert2})). The second vertex is also
present in the Gauge Theory case while the first one pertains to
General Relativity only. In a large \textbf{N} perspective, the
more convenient way to look at the Lagrange multiplier field
$\phi$\ is to consider it as a propagating field with a very high
mass: this point of view allows an interesting interpretation of
its physical role as it will be shown in the last section.
The last term on the right hand side of eq. (\ref{spa}) (defined in eq. (\ref%
{enfo1})) should not be considered as true vertex: it enforces some
restrictions\footnote{%
Such restrictions become less and less important at large \textbf{N:} to see
this, it is enough to note that these restrictions imply that one of the
component of $\phi$ can be expressed in terms of the others. When \textbf{N}
is large (since the number of components of $\phi$ grows as \textbf{N}$^{4}$%
) this fact is not relevant.} on the internal index structure of
$\phi$ and will not be relevant as far as the large \textbf{N}
counting is concerned.

Eventually, the last point to be clarified before to carry on the
large \textbf{N} limit is the scaling of the coupling constant
$\kappa$: in the large \textbf{N} limit, it is natural to assume
that the product
\begin{equation}
\gamma_{e}=\mathbf{N}\kappa,   \label{effcoupl}
\end{equation}
which plays the role of effective coupling constant, as fixed.

\begin{figure} [h]
\begin{center}
\includegraphics*[scale=.50]{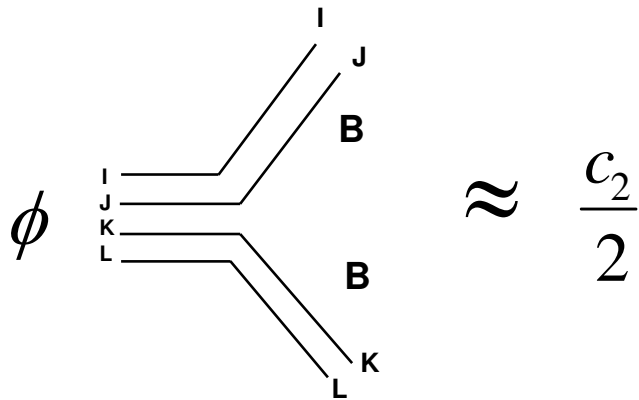}
\end{center}
\caption{This vertex (in which a field represented by four
internal lines appears) only pertains to General Relativity. Such
a field plays a very peculiar role as it will be shown in the
following.} \label{Vvert1}
\end{figure}

\subsection{The General Relativity 't Hooft limit}

Now it is possible to perform the 't Hooft limit: the main object of
interest is the free energy so that only fat graphs with no external legs
will be considered.

\begin{figure}[h]
\begin{center}
\includegraphics*[scale=.45]{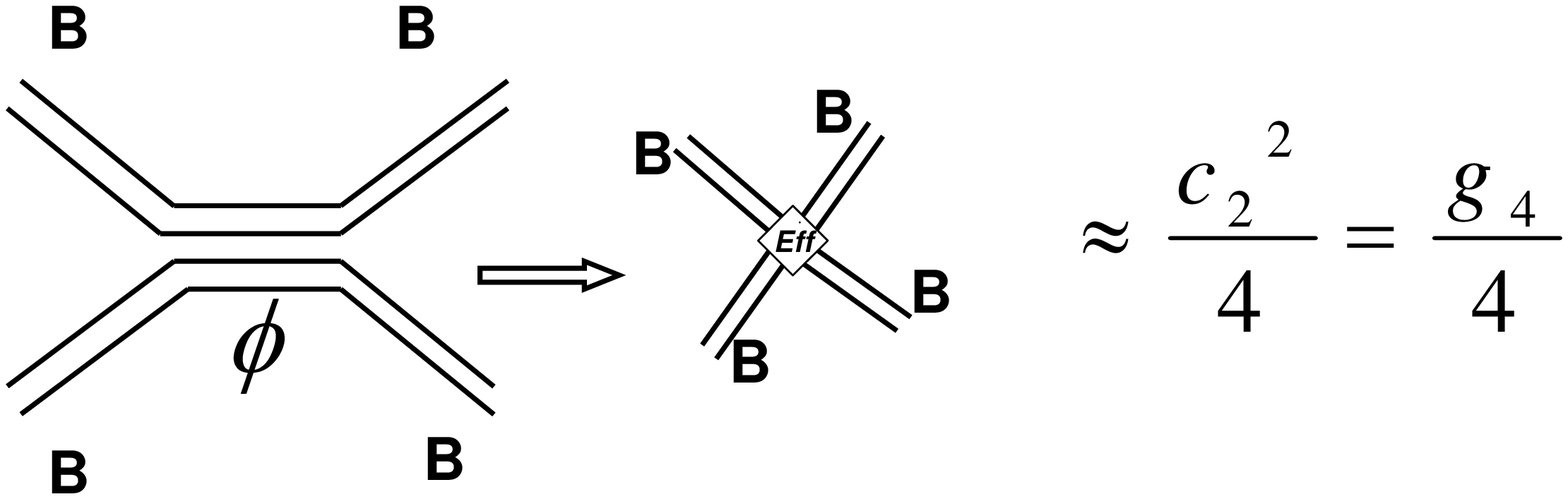}
\end{center}
\caption{The vertex with $\protect\phi$ gives rise to an effective
quadruple vertex for $B$. However, being $\protect\phi$
represented by four internal lines, this quadruple vertex is not
on equal footing with standard gauge theoretical vertices. This
fact has a holographic interpretation.} \label{EffVert}
\end{figure}

However, before to proceed, a point needs to be clarified. The vertex $V_{2}$
in eq. (\ref{vertex}) behaves in some sense (actually, as it will be
explained in the last section, there is an important difference with a non
trivial physical interpretation) as an effective quadruple vertex for $B$
(see fig. (\ref{EffVert})): the point is that the field $\phi$ is only
coupled to $B$ so that, in closed graphs with no external legs the vertices $%
V_{2}$ are always in pairs. Thus, the effective coupling constant $g_{4}$ of
the quadruple vertex is%
\begin{equation}
g_{4}=\left( c_{2}\right) ^{2},   \label{coup1}
\end{equation}
while the coupling constant of the cubic vertex is simply $g_{3}$.

At last, it is possible to apply the standard "large \textbf{N} counting
rules" for fat graphs (see, for two detailed pedagogical reviews, \cite{Di99}%
). These counting rules can be deduced by using \textit{matrix
models}; it is usually convenient to choice the coupling constant
of $n-$uple vertex by dividing it by $n$ \cite{Di99}: this is the
reason behind the normalization in eq. (\ref{vertex}). Thus, the
usual \textit{matrix models} techniques \cite{Di99} tell that a
generic fat graph $\Gamma$ with no external legs will have the
following dependence on \textbf{N} and on the coupling constants:

\begin{equation}
W_{\Gamma}(E,V,n_{p})=\kappa^{E-V}\mathbf{N}^{F}%
{\displaystyle\prod\limits_{p}}
g_{p}^{n_{p}},\quad%
{\displaystyle\sum\limits_{p}}
n_{p}=V \label{factor1}%
\end{equation}

where $E$ is the number of propagators, $n_{p}$ is the number of
$p-$uple vertices in the graph $\Gamma$ (in this case the only
vertices to be counted are the ones with $p=3$; the physical role
of the effective quadruple vertex will be analyzed in the fifth
section) and $F$ is the number of faces of the fat graph. In this
purely gravitational case with no matter
fields with one internal index, one has%
\begin{equation}
F=h   \label{pugra}
\end{equation}
where $h$\ is number of closed "color" loops of $\Gamma$.

By using eq. (\ref{pugra}) and the well known Euler formula%
\begin{equation}
2\overline{g}-2=E-V-F,   \label{eule1}
\end{equation}
where $\overline{g}$ is the least genus\footnote{%
There is a subtlety here in the definition of genus $\overline{g}$
(this point will be discussed in the next sections) related to the
fact that the fundamental representation of the internal gauge
group is real. For this reason the notation $\overline{g}$
(instead of the usual one $g$) will be adopted.} of a Riemann
surface on which the fat graph $\Gamma$ can be drawn without
intersecting lines, the weight $W_{\Gamma}$ of the fat graph turns
out to be

\begin{align}
W_{\Gamma}(E,V,n_{p})  &  =\kappa^{2\overline{g}-2}\mathbf{\gamma}_{e}^{h}%
{\displaystyle\prod\limits_{p}}
g_{p}^{n_{p}} \label{genexp}
\end{align}

where the effective coupling constant $\gamma_{e}$ has been
introduced in eq. (\ref{effcoupl}).

\begin{figure}
\begin{center}
\includegraphics*[scale=.40]{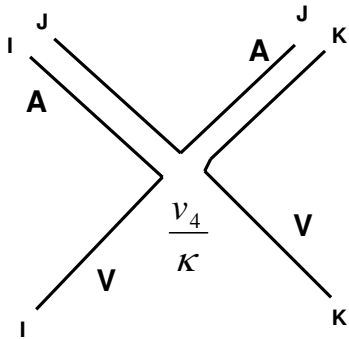}
\end{center}
\caption{This is an example of a planar graph with eight internal
index loops and twelve triple vertices.} \label{ultima10}
\end{figure}

At a first glance, this result gives rise to the usual topological
expansion for the free energy $\mathit{F}$, similar to the one of
the Gauge Theory case, as a sum of the above factor in eq.
(\ref{genexp}) times a suitable space-time-momentum factor
$F_{\Gamma}$ over the closed
connected fat graphs%

\[
\mathit{F}=%
{\displaystyle\sum\limits_{\substack{\Gamma closed\\connected}}}
W_{\Gamma}(E,V,n_{p})F_{\Gamma}%
\]

in which the leading term in the genus expansion is the planar one and the
corrections in the topological expansion are suppressed as powers of $1/%
\mathbf{N}^{2}$. In fact, there are interesting differences, related both to
the gauge group and to the vertices, which will be analyzed in the next
sections.

\section{The inclusion of matter and the Veneziano limit}

In this section the inclusion of matter in the gravitational 't Hooft limit
and the Veneziano limit will be discussed.

\begin{figure}[h]
\begin{center}
\includegraphics*[scale=.40]{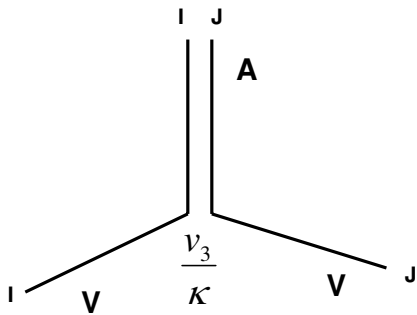}
\end{center}
\caption{This is a triple vertex with two matter fields and a
gravitational connection.} \label{vertmatt11}
\end{figure}

\begin{figure}[h]
\begin{center}
\includegraphics*[scale=.40]{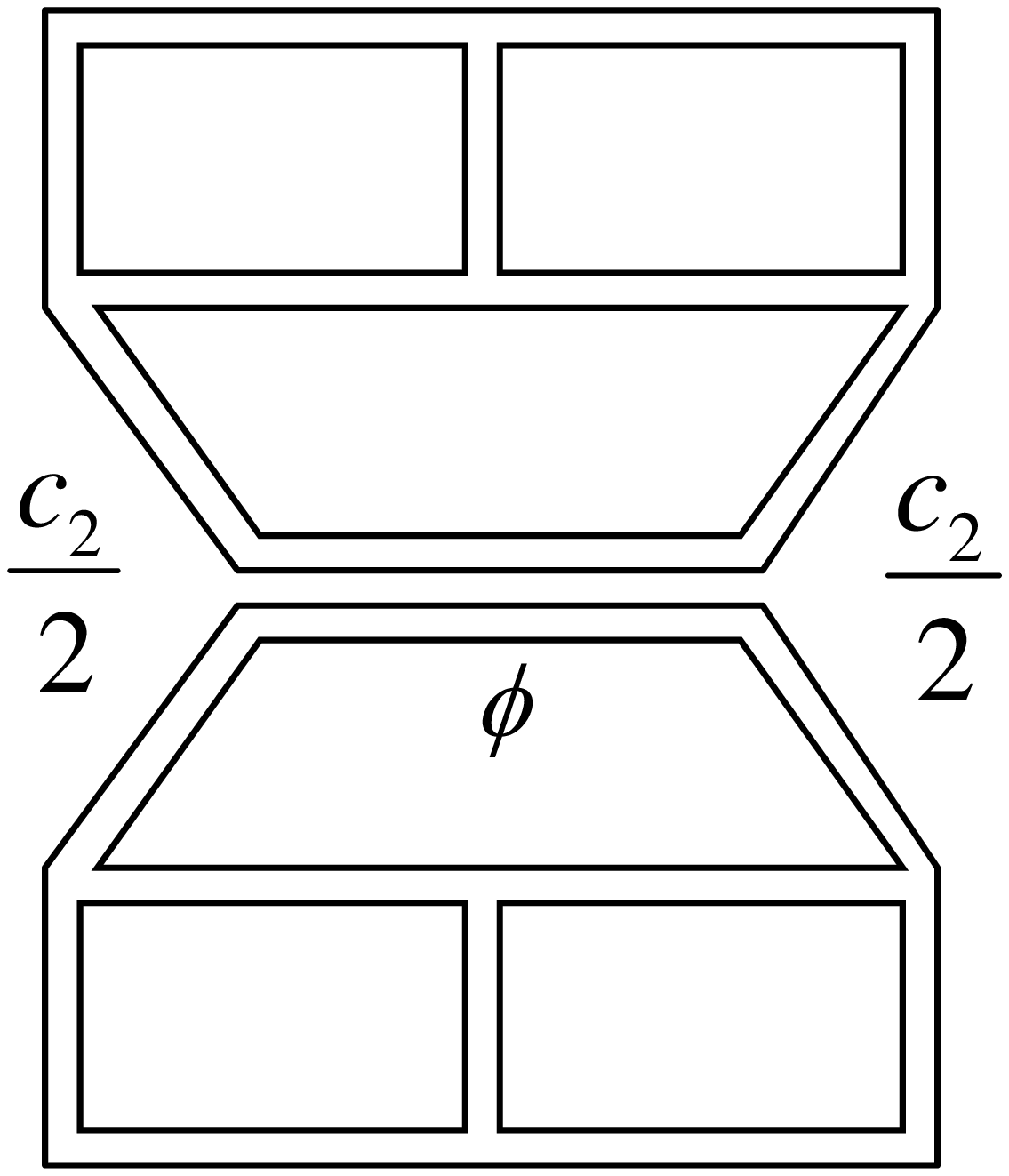}
\end{center}
\caption{This is a quadruple vertex with two matter fields and two
gravitational connections.} \label{vertmatt22}
\end{figure}

Once the purely gravitational 't Hooft limit has been introduced,
the inclusion of matter is the natural further step. However, the
situation is less clear than in the Gauge Theory case. Vectors, in
the standard metric formalism, are coupled to gravity via the
Levi-Civita covariant derivative which, of course, acts on its
vectorial index. Therefore, in this scheme, vectors are
represented as scalar particles with an internal index $J$
running from $1$ to $\mathbf{N}$%
\begin{equation*}
V_{\mu }\rightarrow V_{J}.
\end{equation*}%
and should be coupled to the gravitational connection $A$ by terms $\Upsilon
_{i}$ as (see fig. (\ref{vertmatt11}) and fig. (\ref{vertmatt22}))%
\begin{equation}
\Upsilon _{4}\sim \frac{v_{4}}{4\kappa }A_{\mu
}^{IJ}V_{J}A_{IL}^{\mu }V^{L},\quad \Upsilon _{3}\sim
\frac{v_{3}}{3\kappa }\left( p _{\mu }V^{J}\right) A_{JL}^{\mu
}V^{L}  \label{vecoupl}
\end{equation}%
where $v_{i}$ are coupling constants normalized in a suitable way to take
advantage of the (already mentioned) large \textbf{N} counting techniques
\cite{Di99}. The above vertices could come from, for example, a kinetic term
of the form%
\begin{equation*}
\nabla _{A}V^{J}\nabla _{A}V_{J}
\end{equation*}%
where $\nabla _{A}$ is the covariant derivative of the connection
$A$.

For spinors the situation is more involved: the covariant
derivative $\nabla _{\gamma }$ on a generic spinor $\Psi $ in the
standard metric formalism
reads%
\begin{align*}
i(\gamma ^{\mu }(x)\partial _{\mu }-\gamma ^{\mu }(x)\Gamma _{\mu })\Psi &
=\nabla _{\gamma }\Psi , \\
\gamma ^{\mu }(x)& =e_{K}^{\mu }\gamma ^{K}
\end{align*}%
where $e_{K}^{\mu }$ are the vierbeins, $\Gamma _{\mu }$ is the
spinorial Levi-Civita connection (in which, besides the connection
$A$, also the vierbeins enter: see, for example, \cite{Ch84}) and
$\gamma ^{K}$ are the standard flat Dirac matrices. The problem is
that in the BF formulation of General Relativity the vierbeins do
not appear directly: the fundamental field in the first order BF
formalism is $B$ which, actually, is the exterior product of two
vierbeins. Thus, it is not clear how to construct interaction
vertices with spinors. For this reason, here it will be considered
only the contribution of vectors. It is worth to stress here that,
in any case, the coupling terms with spinor should not be very
different from $\Upsilon _{3}$ and $\Upsilon _{4}$ which,
therefore, provide the role of matter at large \textbf{N} with a
detailed description. This can be argued as follows: a spinor is
always accompanied by a flat Dirac matrix and by a vierbein and in
a spinor current there are always two spinors. Hence, in a spinor
current, there are always two vierbeins which, from an internal
index perspective, are similar to $B$ and, by the way, $B$ has the
same internal index structure of $A$. Consequently, as far as a
large \textbf{N} counting is concerned, a spinorial vertex is well
described by the vertex $\Upsilon _{4}$ in eq. (\ref{vecoupl}).

However, there is an apparent difficulty in dealing with scalar particles.
Ordinary matter couples to the gravitational connection $A$ through a
vectorial internal index. On the other hand, at a first glance scalars do
not couple to the gravitational connection $A$ since, on them, covariant
derivatives coincide with ordinary derivatives. This difficulty is very
similar to the difficulty which one encounters in dealing with baryons in
large \textbf{N} SU(\textbf{N}) (in this case at a first glance, being the
fundamental representation of $so(\mathbf{N}-1,1)$ real, it is not clear
what states could be analogous to mesons): baryons\footnote{%
Baryons are color singlets made of particles with the same sign
under
charge conjugation: therefore, they are not neutral under charge conjugation.%
} in many respects behave as soliton in a large \textbf{N} expansion \cite%
{Wi79}. In particular, this implies that their (relatively large)
masses are of order of an inverse power of the 't Hooft coupling
and their interactions are suppressed by powers of $1/\mathbf{N}$.
On the gravitational side, this seems to suggest that scalar
particles which are not neutral under charge conjugation (such as
the Higgs boson) should have relatively large masses compared to
vectors and spinors. To provide this suggestive analogy with
quantitative supports would require a detailed analysis of the
space-time-momentum dependent parts of the fat graphs: this is out
of the scopes of the present paper. Indeed, this is a direction
worth to be investigated which could be rich of phenomenological
consequences.

Now, it is possible to include matter fields also in the expansion. In
general, when there are vertices with matter fields which, in the 't Hooft
notation, are represented by single lines (as it happens in the present
case), eq. (\ref{pugra}) is modified in this way%
\begin{equation*}
F=h+L
\end{equation*}%
where $L$ is the number of matter loops in the closed connected fat graph.
On the other hand, matter loops do not contribute to the (exponent of the)
power of \textbf{N} of the fat graph since, due to the interactions, the
closed matter loops do not correspond to closed internal index loops.
Consequently, as one should expect, in this case eq. (\ref{genexp}) has to
be modified as follows%

\begin{align}
W_{\Gamma}(E,V,n_{p},L,n_{v})  &
=\kappa^{2\overline{g}-2+L}\mathbf{\gamma}_{e}^{h}%
{\displaystyle\prod\limits_{p}}
g_{p}^{n_{p}}%
{\displaystyle\prod\limits_{i=3,4}}
v_{i}^{n_{i}} \label{matter1}%
\end{align}

where $v_{i}$ are the coupling constants of the matter vertices in
eq. (\ref{vecoupl}) and $n_{i}$ is the number of matter vertices
with coupling constant $v_{i}$. Thus, in the gravitational case
also "ordinary" matter fields are suppressed in the large
\textbf{N} expansion.

\begin{figure}
\begin{center}
\includegraphics*[scale=.50]{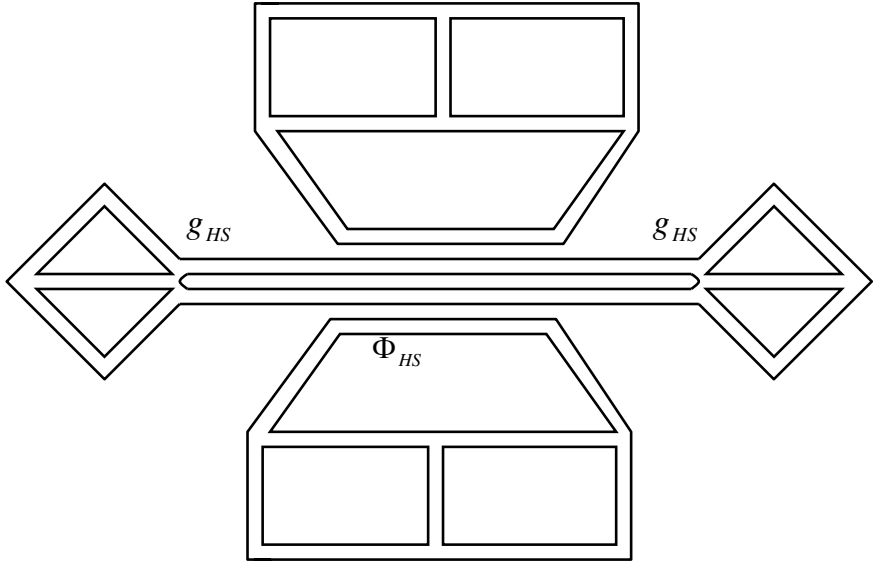}
\end{center}
\caption{Non orientable two dimensional surfaces can be
constructed by gluing N Möebius strips onto a sphere from which N
spherical caps have been removed. Such a surface has genus equal
to N.} \label{moebi1}
\end{figure}

Here it becomes visible a striking difference between the Gauge
Theory and
the General Relativity case. In the purely gluonic sector of large \textbf{N} SU(%
\textbf{N}) Yang-Mills theory, in the topological expansion the
subleading terms are suppressed by powers of $1/\mathbf{N}^{2}$
(in fact, matter loops give rise to factors of the order of powers
of $1/\mathbf{N}$): of course, as it was first discovered by 't
Hooft, this is due to the Euler formula for the genus of
orientable two-dimensional surfaces. In the large \textbf{N}
expansion of SU(\textbf{N}) Gauge Theory only orientable surfaces
enter because the fundamental representation of SU(\textbf{N}) is
not real and the adjoint representation of SU(\textbf{N}) is the
tensor product of the fundamental and the anti-fundamental.
Graphically, this is expressed by adopting \textquotedblright the
arrow\textquotedblright\ notation \cite{T74a} in which the gluon
is represented by two lines having arrows pointing in opposite
directions: this necessarily implies that the fat graph is
orientable. In (the BF formulation of) General Relativity the
situation is different: the gauge group is SO(\textbf{N}-1,1) and
the fundamental representation is real. For this reason, non
orientable two-dimensional surfaces cannot be omitted in the
topological expansion. For non orientable surfaces also there is
an Euler formula which relates the right hand side of eq.
(\ref{eule1}) to the genus of the non orientable surfaces (which
is always a positive integer). Non orientable two-dimensional
surfaces can be obtained by cutting $n$ discs from a sphere and
then attaching $n$ Möebius strips to the sphere by gluing the
boundaries of the Möebius strips with the boundaries of the holes
of the sphere (see fig. (\ref{moebi1})). The surface obtained in
this way is a non orientable surface of genus $g$ equal to $n$.
The Euler formula in this case reads (see, for example, \cite{Ste81})%
\begin{equation}
g-2=E-V-F.   \label{eule2}
\end{equation}
Consequently, when non orientable surfaces are included, the right
hand side of the above equation can be odd as well.

In order to use a unified notation it is more convenient to
consider only eq. (\ref{eule1}) with the convention that
$\overline{g}$ can be both integer (for orientable surfaces) and
half-integer (for non orientable surfaces). Thus, unlike the Gauge
Theory case, in the purely gravitational large \textbf{N}
expansion of General Relativity the subleading terms are
suppressed by powers of $1/\mathbf{N}$ which are of the same order
of matter loops corrections: in a sense, the contributions of non
orientable fat graphs are able to \textquotedblright
mimic\textquotedblright\ matter. This point will be discussed in
slightly more details in the next section.

Another interesting limit worth to be considered in this scheme is
the Veneziano limit. In the Gauge Theory case, the Veneziano limit
\cite{Ve76} had an important role in clarifying non trivial
features of quarks dynamics which in the 't Hooft limit were not
manifest because of the further suppression in $1/\mathbf{N}$ due
to the matter loops. The idea is to keep fixed, in the large
\textbf{N} limit, the ratio $\mathbf{N}_{f}/\mathbf{N}$ (where
$\mathbf{N}_{f}$ is the number of flavour) too: in this way the
suppression due to the matter loops is compensated by a factor $\mathbf{N}%
_{f}$ (of course, we are assuming that the masses of matter fields are the
same otherwise flavour symmetry would be explicitly broken). Consequently,
the weight factor (\ref{matter1}) of the generic closed connected fat graph $%
\Gamma$ with $L$\ matter loops becomes%

\begin{align}
W_{\Gamma}^{V}(E,V,n_{p},L,n_{v})  &
=\left(  \mathbf{N}_{f}\right)  ^{L}\kappa^{2\overline{g}-2+L}%
\mathbf{\gamma}_{e}^{h}%
{\displaystyle\prod\limits_{p}}
g_{p}^{n_{p}}%
{\displaystyle\prod\limits_{i=3,4}}
v_{i}^{n_{i}},\label{altr1}\\
\rho &  =\mathbf{N}_{f}/\mathbf{N.}\nonumber
\end{align}

In this limit, matter loops are not further suppressed: the technical
advantage is that one has at own disposal two natural coupling constants $%
\gamma_{e}$\ and $\rho$ which measure respectively the strength of
the gravitational and of the matter loops.

\begin{figure}
\begin{center}
\includegraphics*[scale=.45]{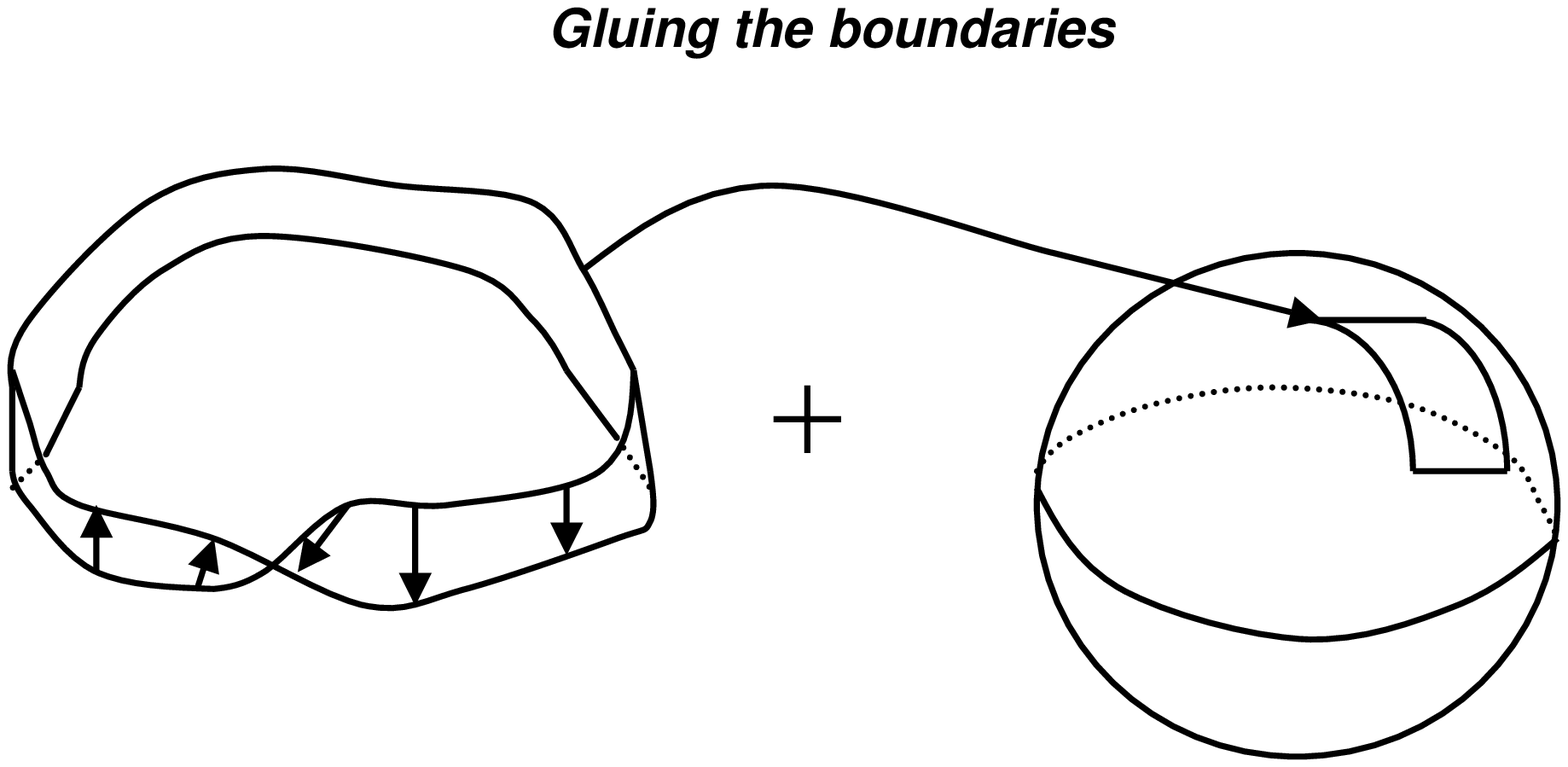}
\end{center}
\caption{This is an example of a planar graph with one matter loop
(represented by the dashed line), four internal index loops, two
triple gravitational vertices and four triple matter vertices.}
\label{ultima11}
\end{figure}

Thus, one can write the following formal expression for the free
energy $\mathit{F}$:

\begin{align}
\mathit{F}  &  =%
{\displaystyle\sum\limits_{\substack{\Gamma closed\\connected}}}
W_{\Gamma}^{V}(E,V,n_{p},L,n_{v})F_{\Gamma}^{V}=\label{energven}\\
&  =%
{\displaystyle\sum\limits_{\overline{g},h,L,n_{p},n_{i}}}
\left(  \mathbf{N}^{2-2\overline{g}}\gamma_{e}^{h}\rho^{L}%
{\displaystyle\prod\limits_{p}}
g_{p}^{n_{p}}%
{\displaystyle\prod\limits_{i=3,4}}
v_{i}^{n_{i}}\right)  F_{\Gamma}^{V} \label{ultife}%
\end{align}

where, as in the previous section, $F_{\Gamma}^{V}$ represents the
spacetime-momentum dependent part of fat graph $\Gamma$ in which also matter
loops and vertices have been included in the large \textbf{N} limit with $%
\rho$ fixed.

\section{Comparing General Relativity and Gauge Theory expansions, Holography and
Higher Spins}

Here some differences between the General Relativity and Gauge Theory large \textbf{N%
} expansions will be discussed and the relation with the Holographic
Principle will be analyzed.

The most evident difference between the two theories manifests itself when
there is no matter: in the purely gluonic sector of large \textbf{N} SU(%
\textbf{N}) the corrections are suppressed by powers of
$1/\mathbf{N}^{2}$ while in the purely gravitational sector of
large \textbf{N} (BF formulation of) General Relativity the
corrections are of order of powers of $1/\mathbf{N}$ (which are of
the same order of matter loops corrections). As it has been
already mentioned, this is due to the contribution of non
orientable fat graphs. Thus, gravity seems to be able to
\textquotedblright imitate\textquotedblright\ matter: this should
not appear really as a surprise. Since the works of Kaluza and
Klein, many purely gravitational higher dimensional models have
been constructed in which gravity in higher dimensions appears in
lower dimensions as gravity plus matter. One could except that for
pure gravity in four dimensions the Kaluza Klein idea does not
provide with matter-like gravitational solutions. In fact, exact
solutions of vacuum four-dimensional Einstein equations which can
be interpreted as spin 1 particles (see \cite{CVV02}) and (more
surprisingly) as spin 1/2 particles (see \cite{FS80a}, for recent
results and an updated list of references see \cite{Ha00}) have
been constructed. The present results tell that this property of
gravity to be able to \textquotedblright look
like\textquotedblright\ matter should survive at a quantum level.

\begin{figure}[h]
\begin{center}
\includegraphics*[scale=.50]{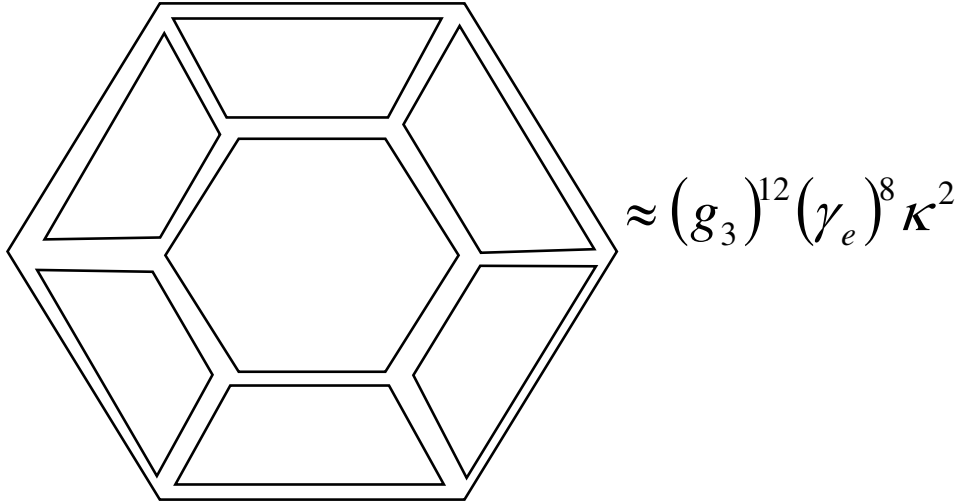}
\end{center}
\caption{This planar graph would appear disconnected into two
pieces without a basic field represented by four internal lines
($\protect\phi$ in this case) which, in fact, makes it connected.
This implies that in theories in which there are fields
represented by more than two internal lines the free energy
receives many more contributions.} \label{gtdisc1}
\end{figure}

There is another difference which is less evident but, perhaps,
more intriguing in a Holographic perspective. As it has been
stressed in the previous sections, the Lagrange field $\phi $
(which, in the double line notation, carries four internal lines)
gives rise to an effective quadruple vertex for the field $B$. In
fact, this effective quadruple vertex is not completely analogous
to a standard quadruple vertex: there is an interesting point
missing in this picture. Let us imagine to give a very large but
not infinite mass to $\phi $ (in other words, we are using a very
powerful \textquotedblright magnifying glass\textquotedblright\ to
disclose the internal structure of the effective quadruple
vertex). It is clear that, to the eyes of a gauge theorists
something strange is happening: many connected fat graphs with
$\phi $ vertices appear as disconnected fat graphs of some more
usual Gauge Theory in which there are not fields represented by
four (or more) color lines. In other words, it is not difficult to
imagine, for example, some \textit{Matrix Model} which, in its
large \textbf{N} expansion, admits these fat graphs: however, this
\textit{Matrix Model} (in which only fields carrying two internal
lines appear) would consider these fat graphs as disconnected and,
therefore, not relevant for computing the free energy. Of course,
in the General Relativity case, these graphs are not disconnected
and \textit{do contribute} to the free energy since $\phi $ is a
basic field of the theory. Thus, in the General Relativity case,
there are many more fat graphs contributing to the free energy
which in a Gauge Theory with fields described by single and double
lines would be neglected (see fig. (\ref{gtdisc1})). The physical
interpretation of this fact could be related to the Holographic
Principle (see, for a detailed review, \cite{Bo02}). The reason is
that, quite generically, since there are
\textquotedblright many more\textquotedblright \footnote{%
Here \textquotedblright many more\textquotedblright\ means
\textquotedblright many more with respect to a gauge theory having the same
fat graphs in the topological expansion but having, in the 't Hooft
notation, only fields represented by single and double
lines.\textquotedblright } terms contributing to the free energy, the free
energy itself is likely to be \textquotedblright higher\textquotedblright
\footnote{%
Here \textquotedblright higher\textquotedblright\ means
\textquotedblright higher than in a gauge theory which has the
same fat graphs in the topological expansion but has only field
represented by single and double lines\textquotedblright .}.

\begin{figure}[h]
\begin{center}
\includegraphics*[scale=.47]{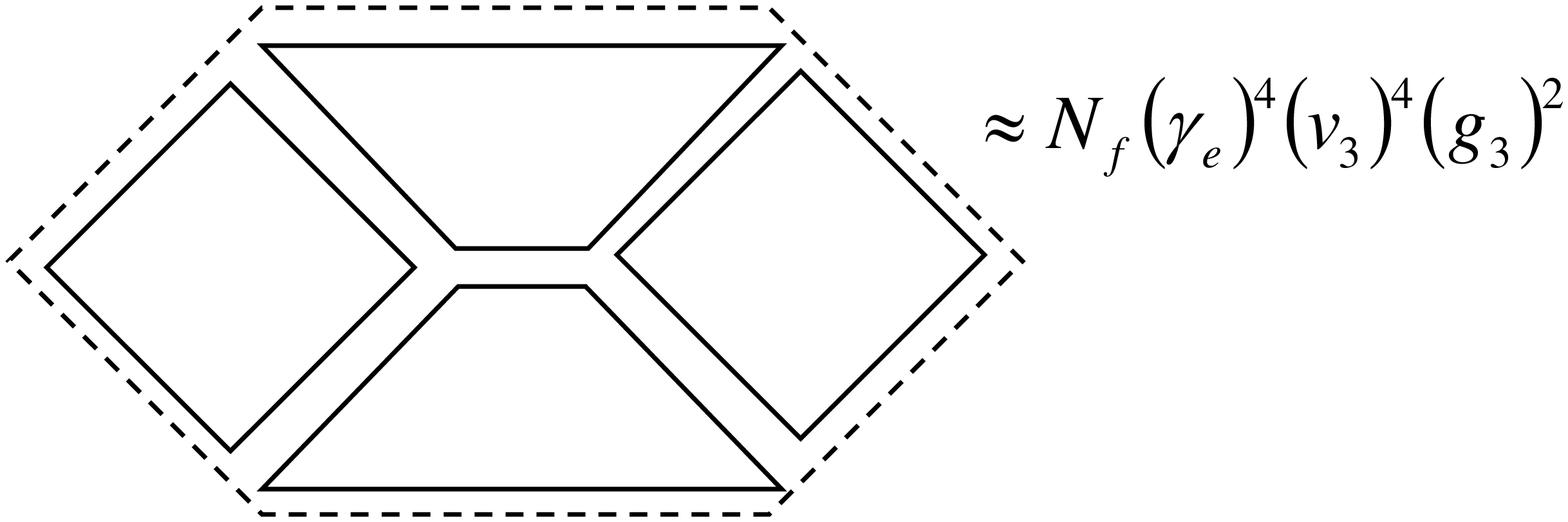}
\end{center}
\caption{In the presence of fields represented by more internal
lines (eight in this case) there are fat graphs which would appear
disconnected (into three pieces in this case) and, in fact, are
connected due to the higher spin field. Here there is an example
of a theory with a field represented by eight internal lines:
similar planar fat graphs give contributions to the free energy
which are absent in ordinary gauge theories.} \label{gtdisc2}
\end{figure}

To provide this last sentence with an analytical proof would
require the analysis of the space-time-momentum dependent part of
a generic fat graph and is a completely hopeless task. However,
there are two quite sound arguments supporting it. Firstly, in
order for the free energy in the General Relativity case not to be
\textquotedblright higher\textquotedblright\ (in the sense
specified above), there should be many fortuitous cancellations in
the sum giving rise to the free energy among terms with different
topological weights $W_{\Gamma }^{V}$. In other words, quite
unlikely, the contribution to the free energy of a given
 \textquotedblright \textit{GT-disconnected}%
\textquotedblright\ fat graph \footnote{%
Which means \textquotedblright Disconnected if interpreted as fat
graphs of a Gauge Theory with only single and double line fields,
but connected when fields represented by more than two internal
lines (such as $\phi $) are taken into account.\textquotedblright
} should be cancelled by the contribution(s) of graph(s) with
different genus, different number of \textquotedblright
color\textquotedblright\ and matter loops and a different
distribution of vertices. The meaning of this fact is
\textquotedblright
Holographic\textquotedblright\ in nature: the free energy can be written as%
\begin{equation*}
\mathit{F}=H-TS
\end{equation*}%
where $H$ is the internal energy, $T$ the temperature and $S$ the
entropy. A \textquotedblright higher\textquotedblright\ free
energy can be seen as a \textquotedblright
lower\textquotedblright\ entropy and this is precisely what one
would expect in a holographic theory: the Holographic Principle
implies a striking reduction of the degrees of freedom (see, for
example, \cite{Bo02}) and, therefore, of the entropy with respect
to a local Quantum
Field Theory \footnote{%
In Quantum Field Theory the entropy, when suitably regolarized, is
proportional to the volume of the space where the fields live.}. The main
role to achieve this decreasing of the entropy has been played by the field $%
\phi $ which, in the 't Hooft notation, is represented by four
internal lines: obviously, the more internal lines are needed to
represent a given field, the more such a field is able to decrease
the entropy because of the many \textquotedblright
\textit{GT-disconnected}\textquotedblright\ fat graphs (see fig.
(\ref{gtdisc2}) in which there is a fat graph with a higher spin
fields $\Phi _{HS}$ represented by eight internal lines
interacting with ordinary fields through the coupling constant
$g_{HS}$).

The second argument supporting this scheme is related to
\textit{string theory}: \textit{string theory} is expected to be a
holographic theory but, unfortunately, is very far from being
solved. However, in \textit{string theory} are predicted an
infinite number of higher spin fields which have very interesting
geometrical properties (see, for example, \cite{BCS04}). Such
fields, in the present notation, would be represented by many
internal lines (according to their spin: the higher the spin, the
more the internal lines). Thus, if there are no fortuitous
cancellations, the entropy of the theory with these higher spin
fields included would be strongly reduced. Hence, higher spin
fields could play the main role in making \textit{string theory}
Holographic. Up to now, a microscopic mechanism able to explain,
at least qualitatively, what kind of interactions could reduce the
entropy as required by the Holographic Principle has not been
found yet. The present results suggest that such a microscopic
mechanism could be related to the interactions of higher spin
fields which, being represented in the present notation by
multiple internal lines, could give rise to the desired reduction
of the entropy. It is interesting to note that this is the first
precise microscopical mechanism which could be able to explain the
Holographic Principle and it is based on higher spin fields which
are very natural objects in \textit{string theory}.

Eventually, it is worth to note the close parallelism between the
BF formulation of General Relativity and the \textit{unfolded}
formulation of higher spin dynamics due to Vasiliev \cite{Va94}:
in both cases, the dynamics is formulated as a trivial
''topological'' dynamics plus a constraint which gives a non
trivial content to the theory. In the \textit{unfolded}
formulation of higher spin dynamics the basic equations can be
reduced to
(see, for detailed reviews, \cite{BCS04})%
\begin{align}
d\omega & =\omega\wedge\omega,\quad\omega=dx^{\nu}\omega_{\nu}^{a}T_{a}
\label{va1} \\
\nabla_{\omega}\overline{B} & =0,\quad\overline{B}=\overline{B}^{A}T_{A}
\label{va2} \\
\chi(\overline{B}) & =0   \label{va3}
\end{align}
where $\omega$ are one forms taking values in some Lie (super)algebra $%
\pounds $\ with generators $T_{a}$, $\overline{B}$ are zero forms taking
values is some (in general) different representation of $\pounds $, $%
\nabla_{\omega}$\ is the covariant derivative associated to $\omega$ and $%
\chi(\overline{B})$ is an algebraic constraint which is invariant under the
gauge transformations of the first two equations (\ref{va1}) and (\ref{va2}%
). If one would neglect eq. (\ref{va3}), then eqs. (\ref{va1}) and (\ref{va2}%
) would be solved by pure gauge fields. Indeed, eqs. (\ref{va1}) and (\ref%
{va2}) bear a strong resemblance with the eq. (\ref{vabf}) of the
BF model, while the few differences appear to be technical in
nature. The main suggestion related to such a close parallelism
between the BF formulation of General Relativity and the
\textit{unfolded} formulation of higher spin dynamics is that the
BF formulation of General Relativity could be very useful to find
a local Lagrangian for interacting higher spin fields.

\section{Conclusions and Perspectives}

In this paper a large \textbf{N} expansion for General Relativity
has been proposed. It is based on the BF formulation of General
Relativity in which the Einstein-Hilbert action is splitted into a
topological term plus a constraint. The scheme proposed allows to
overcome some technical problems present in other proposals - such
as the impossibility to evaluate the exact dependence of a given
fat graph on the small expansion parameter(s). This method allowed
to show that, unlike ordinary Gauge Theory, in the purely
gravitational sector of the theory in the large \textbf{N}
expansion the
subleading terms are of order of powers of 1/\textbf{N }(and not 1/\textbf{N}%
$^{2}$\ as it happens in ordinary Gauge Theory) and so they are of
the same order of matter loops corrections. The technical reason
is that, being the gauge group SO(\textbf{N}-1,1) whose
fundamental representation is real, in the topological expansion
non orientable fat graphs cannot be excluded. This can be related
to the fact that General Relativity is, in a sense, able to
\textquotedblright imitate\textquotedblright\ matter: besides the
well known Kaluza-Klein mechanism, classical exact solutions of
vacuum four dimensional Einstein equations describing spin 1/2 and
spin 1 particles are available too. The present results tell that
such a property should be kept by the theory also at a quantum
level. It is also possible to include matter in this scheme: it
has been stressed that it is not clear how to include scalars in
this picture. At a first glance, it seems that scalars, which have
not \textquotedblright SO(\textbf{N}-1,1)-color\textquotedblright,
could be analogous to baryons in SU(\textbf{N}): this could
explain why they are so heavy (so heavy that they have not been
observed yet) and weakly interacting. Another interesting outcome
of the analysis is the role of fields represented by more than two
internal lines (higher spin fields). The presence of higher spin
fields implies that, quite generically (this means
\textquotedblright unless fortuitous cancellations
occur\textquotedblright) the free energy is higher or,
equivalently, the entropy is lower than in ordinary Gauge Theory.
This could be the microscopical mechanism responsible for the
Holographic Principle which implies a striking reduction of the
degrees of freedom. Moreover, higher spin fields are very natural
objects in string theory. There are many directions worth to be
further analyzed. First of all, it would be very important and
rich of phenomenological consequences (from particles physics to
cosmology) to clarify the nature of scalars in this scheme and, in
particular, if they could be considered as a sort of baryons. A
deeper understanding of the higher spin fields in a Holographic
perspective is also welcome: the dynamics of higher spins, as this
method clarifies, is likely to have a very strong influence on the
microscopical entropy.

\begin{ack}
This work has been partially supported by PRIN SINTESI 2004.
\end{ack}

\end{document}